# Self-Organized Nanorod Arrays for Large-Area Surface-Enhanced Infrared Absorption

M. C. Giordano[a,], M. Tzschoppe[b], M. Barelli[a], J. Vogt[b], C. Huck[b], F. Canepa[a], A. Pucci[b,*],

and F. Buatier de Mongeot[a,*]

([a]) Dipartimento di Fisica, Università di Genova, Via Dodecaneso 33, 16146 Genova, Italy

([b]) Kirchhoff Institute for Physics, University of Heidelberg, Im Neuenheimer Feld 227, 69120 Heidelberg, Germany

ABSTRACT

Capabilities of highly sensitive surface-enhanced infrared absorption (SEIRA) spectroscopy are demonstrated by exploiting large-area templates (cm$^2$) based on self-organized (SO) nanorod antennas. We engineered highly dense arrays of gold nanorod antennas featuring polarization-sensitive localized plasmon resonances, tunable over a broadband near- and mid-infrared (IR) spectrum, in overlap with the so-called "functional group" window. We demonstrate polarization- sensitive SEIRA activity, homogeneous over macroscopic areas and stable in 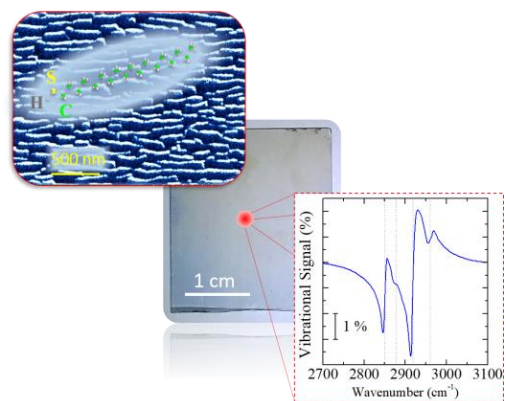 time, by exploiting prototype self-assembled monolayers of IR-active octadecanthiol (ODT) molecules. The strong coupling between the plasmonic excitation and molecular stretching modes gives rise to characteristic Fano resonances in SEIRA. The SO engineering of the active hotspots in the arrays allows us to achieve signal amplitude improved up to 5.7%. This figure is competitive to the response of lithographic nanoantennas and is stable when the optical excitation spot varies from the micro- to macroscale, thus enabling highly sensitive SEIRA spectroscopy with cost-effective nanosensor devices.

KEYWORDS: *plasmonic nanoantennas, self-organized arrays, large-area nanosensors,*

*surface-enhanced infrared absorption (SEIRA), IR spectroscopy, Fano resonances*

# Introduction

Infra-Red (IR) spectroscopy and nanoimaging [1–7] have recently attracted increasing attention, thanks to the possibility to probe vibrational modes of a variety of molecules in a non-destructive way and to detect subwavelength surface polaritons in novel nanomaterials.[8–12] In particular, Fourier Transform IR (FTIR)[13,14] spectroscopy has enabled the direct detection of molecular vibrational modes in the Mid-IR spectrum (3-25 $\mu$m), showing the great potential of this spectroscopy in nanophotonics, [15,16] biology, and life sciences.[4,17] However, the possibility to clearly identify specific molecular features in the so-called *fingerprint spectral region* (700-1500 $cm^{-1}$) is difficult due to the presence of many vibrational bands. Conversely, the molecular vibrations of functional groups with hydrogen, carbon, and nitrogen can be easily identified in the functional group window extending from about 2000 to 3500 $cm^{-1}$. A great advantage is given by the low-energy, nondestructive excitation at near- and mid-IR frequencies. However, the tiny absorption cross section of molecules practically limits the detection sensitivity in standard FTIR spectroscopy and also in grazing-incidence IR reflection−absorption spectroscopy (IRRAS) [18,19] with which vibrational signals of adsorbate monolayers can be detected on continuous metallic films. Strong light−matter interaction of plasmonic nanoantennas [20–26] offers a unique opportunity for the effective improvement of the IR absorption sensitivity below the nanomolar concentration level in surface-enhanced infrared absorption (SEIRA) spectroscopy. [7,27–33] Recently, highly reproducible SEIRA activity has been demonstrated in well-defined noble metal nanoresonators engineered by top-down nanoscale lithography. [27,28,31] Here, the nanoscale control on the nanoantennas size, shape, and/or interspacing has provided a deep insight into more efficient nanoresonator configurations, highlighting the key role of the active hotspots in IR environmental monitoring [34] and biosensing. [7,31,35] In parallel, IR detection has been recently demonstrated by exploiting lithographically designed nanodevices based on plasmonic nanostructures or two-dimensional materials. [7,30,36–38] In this way, high sensitivity and tunable IR optical response can be achieved, but the lithographic nanofabrication practically limits the active sensing area in the range of $100 \times 100$ μm$^2$, demanding for highly focused optical beams and/or complex detection configurations. [13–20]

Conversely, the capability to detect and identify a set of molecules or biomarkers within unknown biological samples is highly desirable for all daily life biomedical applications and has recently motivated the development of advanced microfluidic biosensor chips [39,40] based on nanopatterned plasmonic templates. In this context, the possibility to fabricate reliable, large-area plasmonic platforms, [41–44] which host a high density of IR-active hotspots, represents a crucial issue in view of cost-effective biosensors, potentially relevant for medical diagnostics and monitoring. [45,46] So far, pioneering SEIRA spectroscopy experiments were performed by exploiting large-area nanoparticle films [47,48] endowed with relatively weak IR molecular absorption, in the range of 0.7%, and limited homogeneity over a macroscopic scale.

Here, we demonstrate highly sensitive and reproducible IR biosensing capabilities enabled by self-organized (SO) plasmonic nanorod arrays. Gold nanorods are confined on top of large-area (cm$^2$) nanopatterned glass templates with maskless, single-step nanofabrication. The method enables the engineering of aligned Au nanorod arrays featuring polarization-sensitive broadband localized plasmons tuned in the near- and mid-IR spectrum (2500−5000 cm$^{-1}$), in resonance with the functional group window. Under this condition, we demonstrate superior SEIRA sensitivity at the monolayer level, detecting a homogeneous response over macroscopic sensor areas. The peculiar capability of this SO approach to control the growth of aligned plasmonic nanorods with high density (∼40 μm$^{-2}$) of nanoscale gaps allows us to strongly enhance the SEIRA sensitivity with respect to alternative large-area plasmonic templates characterized by active antenna spread on the surface (density ∼ few $\mu m^{-2}$, [48]). Here, the self-organized engineering of the active hotspots enables the detection of Fano-like lineshapes in SEIRA from self-assembled monolayers of octadecanthiol (ODT) molecules with visibility improved up to 5.7% and high stability over large areas. This figure arises from an enhancement factor of the single SO plasmonic antennas of about 75000, which is competitive to that of lithographic antennas [27,35] confined on microscopic active areas, thus enabling large-area IR spectroscopy and applications in cost-effective broadband on-chip biosensors.

**Results and discussion**

Anisotropic nanoscale wrinkles extending over large cm$^2$ areas are induced on the surface of commercial microscope glass slides by self-organized nanopatterning. [49] The ion-induced wrinkling instability in amorphous substrates is here exploited (see the Methods section) for driving the growth of ordered arrays of high aspect ratio nanowrinkles bound by faceted ridges as long as several micrometers, as shown by the atomic force microscopy (AFM) image in Figure 1a. After the high ion fluencies amounting to $1.4 \times 10^{19}$ ions/cm$^2$, the ion-driven wrinkling instability enables the growth of nanostructures as high as 100 nm with a periodicity of about 250 nm (Figure 1b), which are characterized by an asymmetric faceted profile (typical slopes of the two ripple ridges of 30 and 50°, respectively). [49]

The highly ordered faceted templates allow us to easily confine quasi-1D arrays of tilted noble metal nanostripes by a single-step maskless process based on kinetically controlled noble metal evaporation at a glancing angle. The scanning electron microscopy (SEM) image of Figure 1c shows the example of Au nanostripe arrays confined on the steeper ripple ridges (50−60° with respect to the sample plane). These facets were directly oriented toward the metal-vapor beam during evaporation, performed at a polar incidence angle θ = 70° of the beam with respect to the normal direction to the surface plane (see side- and top-view sketches in Figure 1b,c, respectively). Under these conditions, nucleation of Au nanostripes as long as several micrometers takes place in registry with the ripple ridges that run perpendicularly to the Au beam projection on the surface (Figure 1c,d, sample A).The mean width, $w$, of the Au nanostripes lying on the tilted facets has been extracted by a statistical analysis of the SEM images and reads 100 nm, while the metal thickness, $h$, coating the facets reads 23 nm and has been monitored in situ during growth. [50] The SEM images show conformal growth of laterally disconnected metallic nanostripes on the Au-illuminated ridges of the SO dielectric template, characterized by a high degree of lateral order in the transverse direction (see zoomed-in SEM image in Figure 1d). In parallel, the length of the Au nanostripes is broadly distributed from 1 μm to about 5 μm.

Provided the uniaxial anisotropy of these quasi-1D metallic arrays, a strong optical dichroism has been detected across the whole near-ultraviolet (NUV), visible, and near- and mid-IR spectral regions in extinction, as clearly shown in the spectra in Figure 1e. Here, the normal incidence relative optical transmittance detected for two different polarizations of the incident light, either transversal ($T_\perp$, dashed line, see the sketch in Figure 1c) or longitudinal ($T_\parallel$, continuous line) to the nanostripe long axis, is shown. In the NUV spectrum above 19000 cm$^{-1}$ (wavelength, λ, below 526 nm), the optical response is dominated by the excitation of s−d interband transition in gold. In the VIS spectrum a narrow-band transmittance minimum is selectively detected for transversal polarization (dashed line in Figure 1e) at about 16670 cm$^{-1}$ (i.e., λ = 600 nm) due to the excitation of localized surface plasmon (LSP) resonance along the nanostripe-confined axis; [20,51,52] a high transparency up to 96% up is instead detected in the near- and mid-IR spectral range due to the strong subwavelength confinement in laterally disconnected nanostripes. Conversely, for longitudinal polarization, the transmittance spectrum is dominated by a monotonic decrease within the whole detected IR window below 8000 cm$^{-1}$ due to the stronger Au reflectivity at lower frequencies. Remarkably, the large-area metasurface shows a large linear IR extinction ratio, ER = T$_\perp$/T$_\parallel$, in the range of 4−12 μm, highlighting the potential of these templates as cost-effective polarimetry components in flat optics. [53–55] The detected optical behavior suggests that the plasmonic resonance of stripes as long as several micrometers (Figure 1c,d), when measured parallel to their long axes, is eventually red-shifted below 2000 cm$^{-1}$ beyond the transparency range of the substrate.

Interestingly, following top-down lithographic approaches, [22,27,28,32] the near-field confinement in plasmonic nanoantennas can be exploited for selectively amplifying vibrational modes from IR-active molecular layers that are tuned in the frequency range of the functional group window (i.e., 2000−3500 cm$^{-1}$). To blue shift the plasmonic excitation in the relevant spectral range, the SO method has been modified for engineering the shape of the metallic nanoantennas, with the aim to decrease the characteristic length of the plasmonic nanostripes down to the sub-micrometer range.

By tailoring the ion beam irradiation of the glass substrate at the early stages of the process (i.e., reducing exposure time down to 300 s, fluence ∼2.3 × 10$^{18}$ ions cm$^{-2}$), we demonstrate the possibility to reduce the length of the wrinkle structures, preserving their anisotropic alignment (Figure 2, and cross-sectional profile in Figure 2a,b), thanks to the occurrence of the ion-induced wrinkling instability. [49] The peculiar presence of faceted ridges becomes obvious in the cross-sectional profile extracted along the y-axis (see Figure 2a,b) and in the histogram of local slopes, $\alpha = \tan^{-1}(\partial z/\partial y)$, shown in Figure 2c, which is characterized by a bimodal distribution peaked at ±30°. The AFM topography of Figure 2a (sample B) clearly shows shorter ripple ridges with respect to Figure 1a (sample A) due to the modified nanopatterning conditions. Indeed, in Figure 2a, the length of the ripple ridges is broadly distributed from about 200 nm to about 1 µm. As a comparison, for the much higher ion doses of $1.4 \times 10^{19}$ ions cm$^{-2}$ employed in sample A, the coalescence of the short ripples into elongated faceted ridges extending up to several micrometers has taken place.

In analogy to sample A (Figure 1), the peculiar faceted morphology of the SO wrinkled glass template enables the confinement of tilted Au nanorod antennas in registry with the dielectric facets (SEM image of sample B in Figure 2d). However, here, the glancing angle Au evaporation has been optimized to fully exploit the presence of inter-ripple defects favoring the longitudinal disconnection of the Au nanostripes. In particular, we selected an out-of-plane polar angle $\theta = 80°$ and an in-plane azimuthal angle $\varphi = 30°$ of the vapor beam (see side- and top-view sketches in Figure 2b,d respectively). Under this condition, we demonstrate the capability to control longitudinal disconnections of the Au nanostripes tailored by the SO template at the micrometer and sub-micrometer levels. The SEM image of Figure 2d clearly shows both laterally and longitudinally disconnected Au nanorod antennas supported by the faceted glass pattern. The local thickness, *h*, of the nanorods has been kept fixed at 23 nm, as in the case of Figure 1, while the change of the nanopatterned template and the metal deposition angle has induced a decrease of the mean nanorods width, *w*, to about 60 nm. The distribution of length, *L*, of the Au nanorods (Figure 2e), obtained from statistics of SEM images of sample B, clearly highlights the presence of Au nanorods whose lengths extend from about 100 nm to about 1.3 µm. This figure proves the capability to strongly tailor the nanostripe length by the SO method proposed here; indeed, a strong reduction of the length of characteristic nanorods from several micrometers (sample A in Figure 1c,d) to the sub-micrometer range (sample B in Figure 2d) has been achieved.

Remarkably, a key feature of this SO templates in view of cost-effective biosensing applications is the homogeneity of the nanorods' morphology and the periodic order over large-area samples typically extending up to 2 × 3 cm$^2$ (see the inset in Figure 2e). It should be noticed that the SO nanofabrication approach employed here is different from glancing angle deposition (GLAD). SO exploits geometrical shadowing of the rippled glass template to obtain conformal decoration of the ridges with Au nanorods at a comparatively low metal coverage (∼20 nm), whereas GLAD exploits self-shadowing from initial metallic nanostructures and columnar growth (Zeno effect) under grazing metal deposition conditions up to very high coverages exceeding many hundred nanometers. Typically, GLAD produces nanorods that protrude out from the surface plane, while the SO nanorods are elongated parallel to the substrate plane. Metallic rods from GLAD are beneficial for surface-enhanced Raman scattering (SERS) because their horizontal plasmonic resonance [22] can be excited by visible light at normal incidence. This study on SO gold nanorods instead exploits longitudinal plasmonic resonance for SEIRA sensing with IR light at normal incidence.

In sample B, the produced length distribution of the Au nanorods within the sub-micrometer range leads to the excitation of a broad plasmonic resonance for the longitudinal polarization of the incident light (T$_∥$, continuous line in Figures 3a and SI1 for the complete NUV−Vis−IR spectra). In parallel, the anisotropy of the Au nanorod array induces the strong dichroism in the optical behavior, as shown in the complete NUV−Vis−IR spectra in the Supporting Information (Figure SI1a). Notice the blue shift of this plasmonic resonance and thus the better match to the relevant functional group spectral window in comparison with the spectrum of long nanostripes in sample A (Figure 1e). The wide length distribution and the dense packing of the nanorods in sample B (Figure 2e) induce a plasmonic resonance extending from the frequency of 2500 cm$^{-1}$ to about 5000 cm$^{-1}$. The spectral width is attributed to homogeneous and inhomogeneous broadening, i.e., the incoherent superposition of the extinction signals from the array of nanorods characterized by a broad length distribution and the near-field coupling between antennas. [56]

Electromagnetic finite-difference time-domain (FDTD) simulations (see Supporting Information) allow us to determine the characteristic size of antennas that are resonant in the functional group window, within a simplifying model that considers antennas as half-cylinder structures. [35] Under this assumption, we obtain that Au nanoantennas that are about 760 nm long and supported on a glass substrate have a longitudinal plasmonic resonance in the ODT stretching vibration range (i.e., 2800−3000 cm$^{-1}$). The red shift of the whole broadband plasmonic resonance with respect to the simulation of a single ideal half-cylinder (with the same aspect ratio $h/w \sim 0.4$) is mainly a consequence of the flatter geometry and interantenna coupling in the array. [20,56,57] The resonant antennas coupled to the ODT vibrational modes in this study are thus shorter compared to the values for decoupled lithographic antennas with a cuboid like shape. [32]

The broadband plasmonic near-field enhancement on the surface of the nanorod arrays, originated by the length distribution in Figure 2e, can be exploited for selectively amplifying the molecular signal from a self-assembled monolayer of ODT on the surface (see the Methods section for sample preparation). Indeed, the longitudinal optical transmittance $T_\parallel$, is strongly modulated by the presence of active molecules in the spectral range between 2800 and 3000 cm$^{-1}$, where the vibrational modes of ODT are expected. [27] Conversely, the molecular absorption is negligible for transversal polarization transmittance $T_\perp$. It can be attributed to the small fraction of Au nanorods detached during the molecular coating process (see the SEM image obtained after coating in the Supporting Information, Figure SI2). Prospectively, this minor effect could be easily avoided by fixing the plasmonic nanostructures on the substrates with a nanometric adhesion layer (e.g.,

Cr, Ti), thus further improving the efficiency and stability of the nanorod templates. Indeed, a nanometer thick Ti layer would improve both the adhesion and the uniformity of Au nanorods, avoiding the growth of isolated nanoclusters in the proximity of the nanoantennas. In parallel, a small red shift of the plasmonic resonance with an improved quality factor is expected due to the change of the Au interface. [58,59] The dichroic optical response allows us to easily normalize the detected signal from the molecules by the nonresonant polarization as a reference without the need of a different reference sample. The relative transmittance [$T_\parallel/T_\perp$] spectrum can be calculated from the detected transmittance spectra $T_\parallel$ and $T_\perp$. This allows us to extract the bare vibrational signal as the difference between the relative transmittance detected in the presence of the ODT molecular monolayer, [$T_\parallel/T_\perp$]$_{mol}$, and the optical background from the bare plasmonic arrays, [$T_\parallel/T_\perp$]$_{background}$, extracted as a fit from the detected spectra. Under this condition, we obtain the vibrational spectrum of Figure 3b. This shows multiple asymmetric resonances with Fano-type line shapes. In particular, two modes are detected at the wavenumbers of 2850 and 2918 cm$^{-1}$, which, respectively, correspond to the energy of the symmetric and antisymmetric stretching modes of the CH2 functional groups of the ODT molecule. The amplitude of the detected SEIRA signal is estimated as the difference between the minimum to maximum transmittance of the Fano-type signal. Remarkably, the SO plasmonic template enables the detection of SEIRA signals with signals of 1.3% and 2.8% for the symmetric and antisymmetric vibrational modes, respectively, which is competitive to the typical signals detected by lithographic nanorod antennas. [27,35] Additionally, the strong plasmonic enhancement induced by SO nanorods allows us to detect the weaker symmetric and antisymmetric stretching modes of the CH$_3$ group in ODT (wavenumbers of 2877 and 2960 cm$^{-1}$, respectively).

The Fano-type coupling between the vibrational dipoles of the molecules and the plasmonic near-field at the hotspots of the nanorod antennas leads here to the antisymmetric enhancement of the molecular vibrational modes. [27,60] In parallel, the detuning of the resonantly coupled excitations, i.e., of the narrow-band vibrational mode and the broadband plasmonic resonance, induces the characteristic asymmetry of the modes, which is well explained by the Fano theory.[61] The latter describes the coupling of the modes using the lineshape function

$$f(\varepsilon) = (q + \varepsilon)^2/(1 + \varepsilon^2) \qquad (1)$$

for the extinction signal with $\varepsilon = 2(\omega - \omega_{vib})/\Gamma$, where $\omega_{vib}$ corresponds to the resonance frequency of the vibrational excitation, $q$ corresponds to the asymmetry parameter, and $\Gamma$ corresponds to the vibrational linewidth. By this lineshape, the fit of the experimental data has been performed (see the Supporting Information, Figure SI4), obtaining a characteristic asymmetry parameter $q = -1.5$, which accounts for the coupling and phase shift between the vibrational and plasmonic modes.

The vibrational SEIRA signal ($V_{SEIRA}$) can be evaluated in terms of the enhancement factor, EF, with respect to the signal detected in IRRAS ($V_{IRRAS}$) from an ODT monolayer coating a gold surface in IRRAS (i.e., the incidence angle of the light corresponding to $\beta = 83°$ [19]) as

$$\text{EF} = \frac{V_{SEIRA}}{V_{IRRAS}} \cdot \frac{1}{C} \cdot \frac{2\sin^2(\beta)}{\cos(\beta)} \cdot (n_s + 1) \qquad (2)$$

where $n_s = 1.5$ is the refractive index of the substrate and $C$ is basically the ratio of the average ODT area densities in the focal spot. In IRRAS, a full monolayer is assumed. ODT does not chemisorb on the glass substrate. We assume full monolayers only on the gold nanostructure surfaces of the arrays. Thus, based on the SEM statistics of our gold antenna array, we estimate $C \cong 15\%$ by conservatively considering all of the Au nanorods with length $L \geq 200$ nm as active hotspots with homogeneous SEIRA activity. However, it has been cleared [27,29,35] that the measured SEIRA signal arises almost only from the molecules that are located at the hotspots of the nanoantennas, in correspondence to their ends. [27] To improve our strong overestimate of the SEIRA-active areas, we evaluate the tips area by considering the 10 nm long apexes of each nanorod via SEM statistics (Figure 2d), corresponding to an effective surface coverage $C_{eff} \cong 0.6\%$. Under this assumption, we found that the characteristic enhancement factor of the resonant nanorod antennas (sample B) reads $EF_{eff} \cong 55000$. We stress that this figure is competitive with the efficiency of lithographic nanoantenna arrays, [62,63] even though $EF_{eff}$ is here underestimated. Indeed, in the evaluation of $C_{eff}$, we included all of the families of nanorods present in the SEM statistics (see the histogram of nanorod length in Figure 2e), even if a portion of them are strongly detuned from the molecular resonance and can be considered as nearly inactive.

To further improve the nanorod array sensitivity for biosensing, the vibrational response of the ODT monolayer has been observed on a third sample (sample C) where the density of active hotspot sites was increased by slightly reducing the ion fluence in the nanopatterning process (see SEM characterization of the sample in the Supporting Information, Figure SI3b). The detected optical transmittance and vibrational signals are shown in Figure 3c,d, respectively. A more pronounced dichroic optical behavior is observed in the transmittance spectra (Figure 3c). In analogy to sample B, a strong vibrational response is detected (blue line in Figure 3d), which is characterized by a Fano lineshape of the modes with asymmetry factor $q = -1.3$. This is very similar to the value obtained for sample B (i.e., $q = -1.5$), demonstrating that the broadband nature of the plasmonic resonance allows us to preserve the phase delay between the plasmonic excitation and the vibrational stretching modes, despite the slight change of the distribution of antennas length and/or shape. This is a remarkable feature uniquely provided by the SO nanorods and allows us to exploit these templates to perform large-area vibrational spectroscopy at a constant phase delay.

In parallel, sample C shows a further enhancement of all of the ODT vibrational modes, with amplitude of the SEIRA signal for the $CH_2$ antisymmetric stretching mode (wavenumber of 2918 cm$^{-1}$) as high as 5.7%.

Since the density of antennas longer than 200 nm is here increased by 50% with respect to sample B (see comparison in the Supporting Information, Figure SI3), we estimate for sample C an effective coverage corresponding to the active tip area $C_{eff} \cong 0.9\%$ and, in turn, to an enhancement factor $EF_{eff} \cong 75000$. This figure improved with respect to the case of sample B and demonstrates the superior performances of the these SO templates with respect to state-of-the-art lithographic nanoantenna arrays. [62,63] This result highlights the importance of nonlinear amplification of the vibrational signal when increasing the density of the active sites, which can be attributed to the strong near-field localization in nanoscale interantenna gaps.

The homogeneity of the SEIRA enhancement over a macroscopic area has been proved by probing the molecular monolayer with a conventional large spot IR spectrometer (diameter of the optical spot of about 3 mm). The vibrational signal detected under large-area probing of the ODT monolayer (Figure 3d, black line) clearly shows full quantitative agreement with respect to the microspectroscopic measurement (Figure 3d, blue line), demonstrating the high degree of homogeneity of the plasmonic antennas arrays. The highly reproducible and stable Fano lineshapes in SEIRA highlight the strong potential of these SO nanorod antenna arrays for large-area amplitude and phase-sensitive IR spectroscopy and biosensing.

## Conclusions

We have demonstrated the cost-effective engineering of self-organized nanorod antennas featuring dichroic plasmonic functionalities in the near- and mid-IR spectral range. A novel SO nanopatterning approach combining ion-induced wrinkling in glasses with glancing angle metal evaporation has been developed for tailoring the morphology and optical response of thin Au nanorod antennas. The polarization sensitive excitation of a broadband LSP resonance in the near-IR enables the highly sensitive detection of molecular monolayers with IR active modes overlapping with the low frequency tail of the plasmonic resonance. Under this condition, highly sensitive SEIRA spectroscopy has been performed on self-assembled monolayers of ODT molecules coating the plasmonic templates. The high density and efficiency of active hotspots on the SO nanorod arrays allows us to detect homogeneous Fano lineshapes with amplitude reaching up to 5.7% and enhancement factors per antenna in the range of $10^4-10^5$, demonstrating improved performances with respect to state-of-the-art lithographic nanoantennas. Remarkably, the broadband nature of the plasmonic resonance ensures a high stability of the Fano lineshape even on different samples, thus enabling large-area amplitude- and phasesensitive IR spectroscopy, multiplexed biosensing, and cost-effective biomedical applications.

## Methods

**Glass Nanopatterning**. Standard microscope glass slides (soda lime glass) are heated up to 685 K, near their glass transition temperature, and irradiated with a defocused ion source (Ar+) of energy of 800 eV at the incidence angle $\theta = 30°$ with respect to the normal direction to the surface. During the ion irradiation process, the glass surface is kept electrically neutral by exploiting the thermoionic electron emission from a tungsten filament negatively biased with respect to the sample at $V_{bias} = -13$ V. Under this condition, highly ordered faceted nanopatterns can be achieved when the ion irradiation is prolonged for 1800 s (i.e., ion fluence of $1.4 \times 10^{19}$ ions cm$^{-2}$), as shown by AFM topography in Figure 1a,b.

**Scanning Electron Microscopy (SEM) Imaging and Statistics.** The SEM images have been acquired with a Hitachi column (SU3500) in the back scattering condition with an acceleration voltage of the electron beam of 10 kV. Statistics of SEM images has been acquired to calculate the histogram shown in Figure 2e. The corresponding SEM images acquired within an area comparable to the micrometric optical spots (i.e., in the range of few 100 μm$^2$) and comparable morphology have been obtained by performing SEM imaging within a much larger area of about 5 mm$^2$. In particular, for what concern the histogram of Figure 2e, five SEM images have been acquired at the center, top, bottom, left, and right edges of a circular area with 50 μm radius, respectively. Each SEM image area corresponds to 9872 $\mu$m$^2$, for a total probed area equal to 49.36 $\mu$m$^2$. Antenna lengths are measured using ImageJ software (open access). By means of the "Analyze Particles" function, a contrast threshold is set to recognize and isolate the Au antennas from the SEM image background. The Feret diameter (that quantifies the length between the two most distant points of an object) of each antenna is thus measured, and the counts over the five SEM images are cumulated to create the statistics shown in Figure 2e.

**Sample Preparation for SEIRA Measurements.** The Au nanorod arrays have been treated with oxygen plasma at 150 W at 0.4 mbar for 3 min to provide a clean gold surface for sulfur binding of the octadecanthiol (ODT, $CH_3(CH_2)_{16}CH_2SH$) molecules. According to a literature recipe, the samples were exposed to a 1 mM solution of ODT (Sigma-Aldrich, purity: 98%) in ethanol for exactly 24 h. [19,33] Under this condition, the ODT/Au sulfur bonding promotes the formation of a SAM monolayer on top of the nanorod arrays. To avoid unbound ODT on the surface, we carefully rinsed the samples afterward and dried them with nitrogen gas. This procedure results in a uniform coating with a thickness of around 2.8 nm [33] of the IR-active ODT, which in principle is below the detection limit for an IR transmittance measurement.[47] To provide optimal measurement conditions, resulting in flat baselines, the references (glass substrates) were treated in the same way as the samples. By doing so, influences due to the solvent or the plasma cleaning procedure on the substrate cancel out in the measurements.

**Microscopic and Standard IR Spectroscopy.** The microscopic IR measurements have been performed by a Bruker Hyperion 1000 IR microscope, coupled to a Bruker Tensor 27 FTIR spectrometer in a transmittance geometry. A polarizer within the beam path allows us to align the electric field component of the incoming IR radiation either along the nanorods' long axis (parallel) or perpendicular to it. To avoid disturbing absorptions of water vapor or carbon dioxide in the measurement, the whole beam path was purged with dry air. The signal was detected by a mercury cadmium telluride (MCT) detector that was kept at liquid nitrogen (LN2) temperature. We used a spectral resolution of 2 cm$^{-1}$ and an aperture size of 104.2 $\mu$m in diameter for the measurements. Each spectrum is averaged over several hundred scans to provide a high signal-to-noise ratio, allowing the careful analysis of the molecular ODT vibrational signature. To obtain information from the gold nanorods only, relative measurements were performed, which means that the measurements of the structured sample were referenced to the measurements of the reference glass substrate. IR spectroscopic measurements, using an aperture with a diameter of 3.0 mm, were performed using a Bruker IFS66v/S spectrometer in the transmittance geometry. In this case, the whole beam path was evacuated to 5 mbar to avoid atmospheric distortions. For the measurements, a resolution of 2 cm$^{-1}$ and either perpendicular or parallel polarized light were chosen. Several hundred scans were averaged to obtain a high signal-to-noise ratio, measured with an MCT detector kept at LN$_2$ temperature. As already explained for the microscopic measurements, relative measurements have been performed.

**FIGURES**

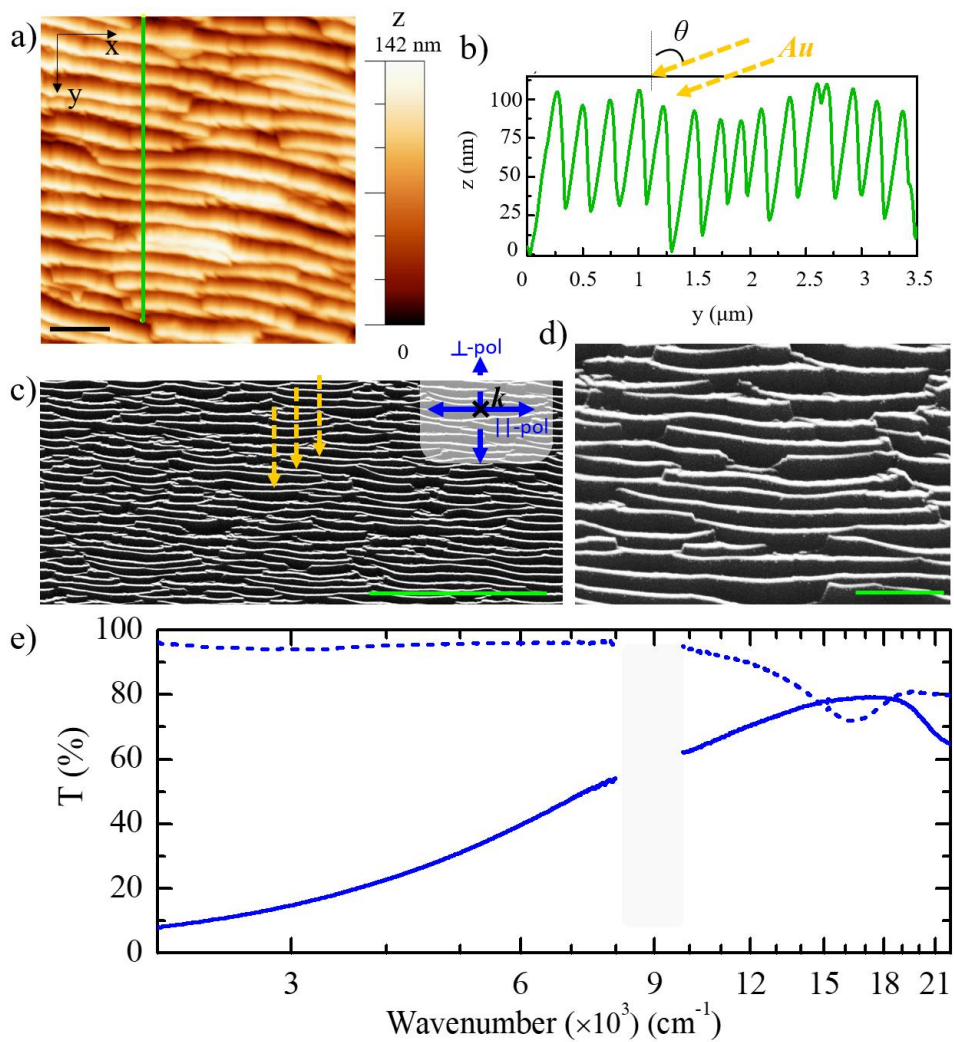

**Figure 1. a,b)** AFM topography and cross section profile, respectively, of a high aspect ratio rippled glass template. The scale bar in panel a) corresponds to 800 nm. **c,d)** Large area and zoomed SEM image of quasi-1D Au nanostripe arrays confined on the self-organized glass

template. The scale bars in c) and d) correspond to 5 μm and 1 μm, respectively. **e)** Relative optical transmittance (T) spectra of the Au nanostripe arrays detected for longitudinal (continuous line) and transversal (dashed line) polarization of the incident light with respect to the nanostripes axis. The spectra characterize the response of the arrays from the mid-IR to the near-UV spectral region.

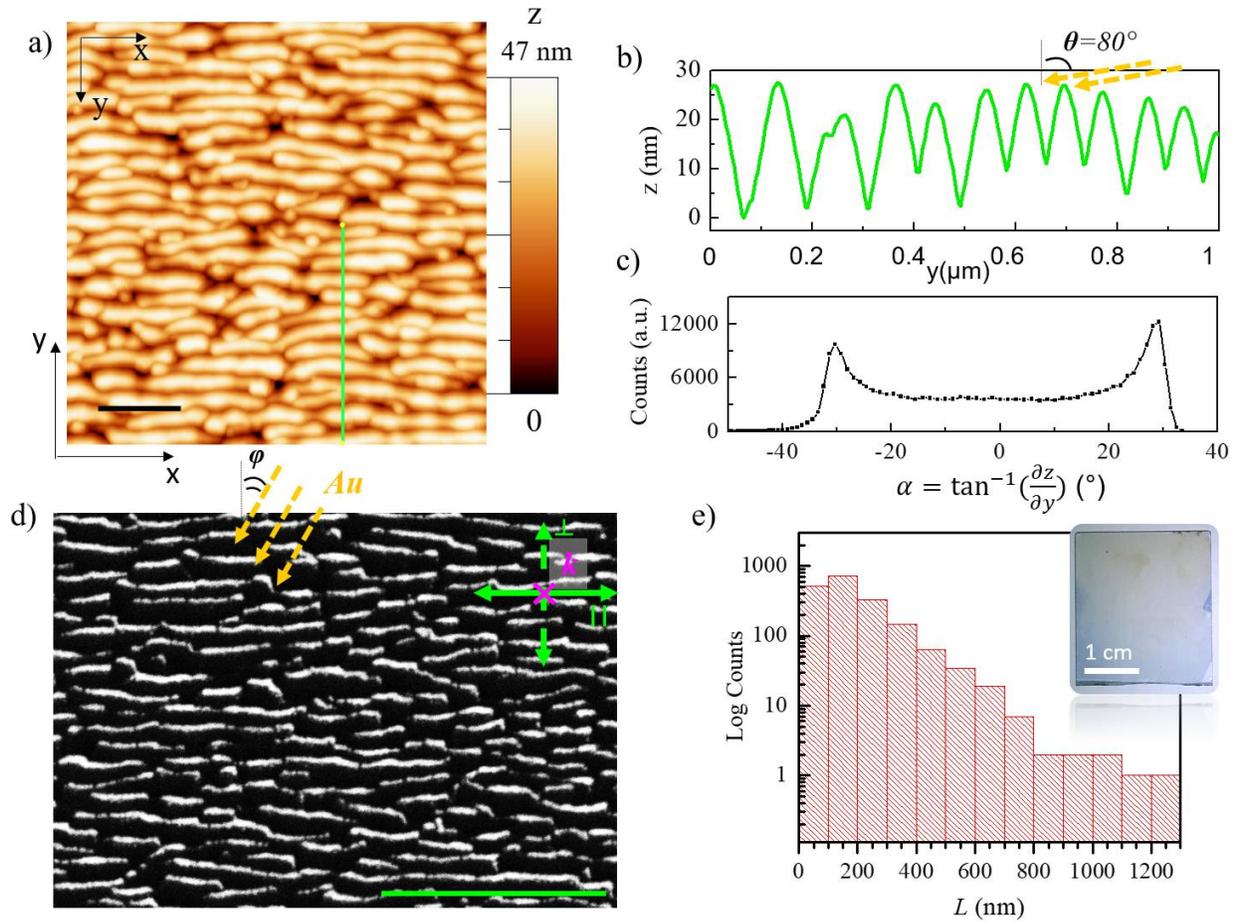

**Figure 2. a,b)** AFM image and cross section profile, respectively, of aligned glass nanorods engineered by defocused ion beam irradiation. The black scale bar in panel a) corresponds to 400 nm. **c)** Histogram of the characteristic lateral slope, $\alpha = \tan^{-1}\left(\partial z / \partial y\right)$, of the glass nanorods pattern extracted from the AFM image. **d)** SEM image the Au nanorod arrays confined on the glass

template. The blue scale bar in d) corresponds to 1 $\mu$m. **e)** Histogram of the Au nanorod length extracted from a statistic of SEM images acquired over an area of the order of few mm$^2$.

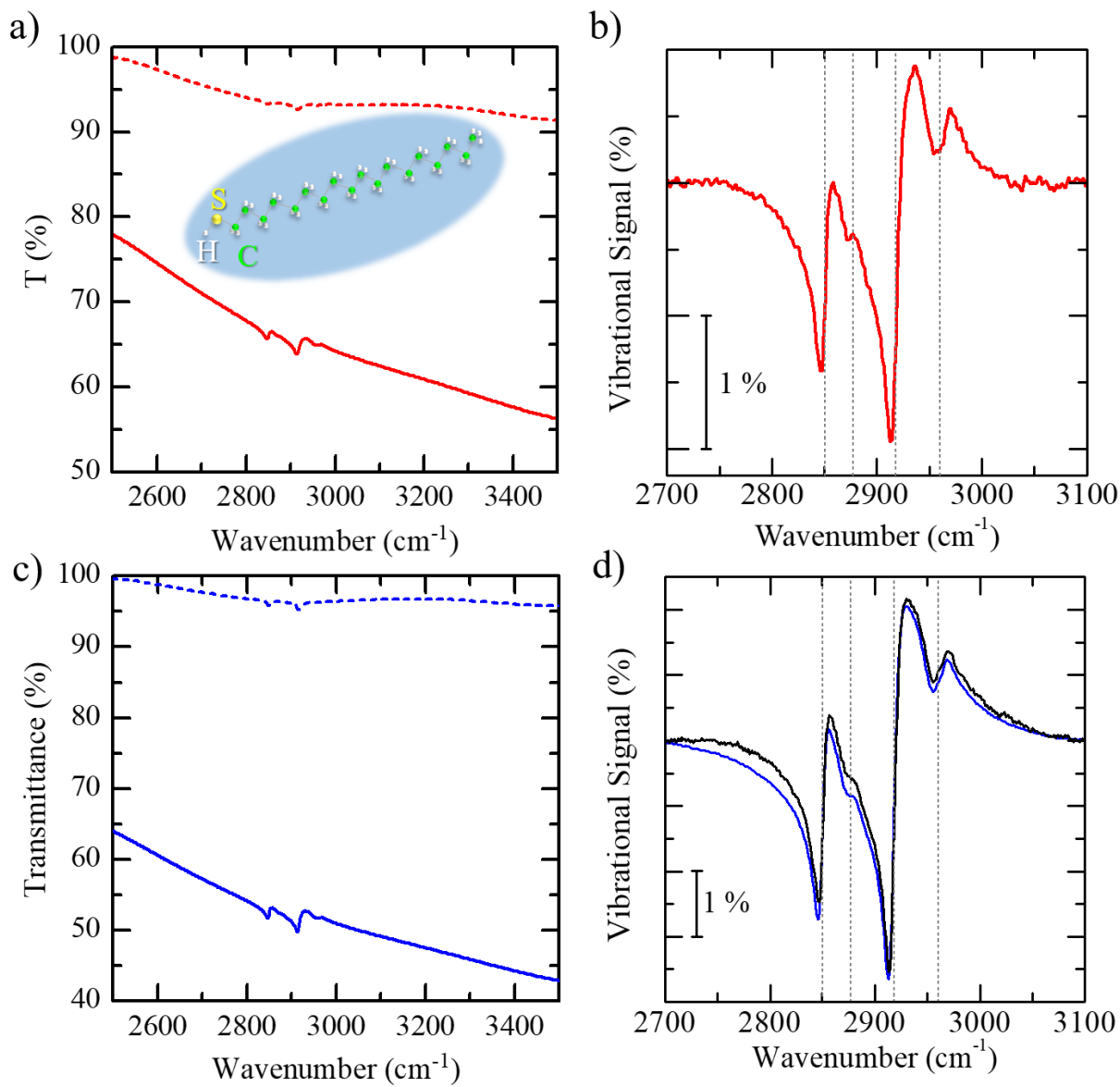

**Figure 3 a,c)** Mid-IR relative optical transmittance (T) spectra of two different samples of Au nanorod arrays functionalized with a self-assembled monolayer of ODT molecules (Sample B and Sample C) which are characterized by increasing density of active hot spots. The continuous and dashed lines spectra refer to longitudinal ($T_{\parallel}$) and transversal ($T_{\perp}$) polarization of the incident light, respectively. **b,d)** Vibrational signal from the ODT monolayer respectively extracted from the two spectra of Fig. 3a, 3b as $[T_{||}/T_\perp]_{\text{detected}} - [T_{||}/T_\perp]_{\text{background}}$. The SEIRA spectra show the characteristic vibrational modes of the ODT molecule which are peaked at 2850 cm$^{-1}$, 2918 cm$^{-1}$ (i.e. symmetric and antisymmetric stretching mode of the functional group CH$_2$) and at 2877 cm$^{-1}$, 2960 cm$^{-1}$ (i.e. symmetric and antisymmetric stretching mode of the functional group CH$_3$). In panel **d)** the vibrational signal detected with the IR optical microscope (blue line – optical spot of 100 μm size) is compared with the signal detected with a standard IR spectrometer (black line - optical spot of 3 mm size).

**Associated Content**: Supporting Information

**Corresponding Authors**

*Francesco Buatier de Mongeot: buatier@fisica.unige.it

*Annemarie Pucci: pucci@kip.uni-heidelberg.de

**Acknowledgements**

F.B.de.M. and M.C.G. thank Ennio Vigo and Roberto Chittofrati for providing technical support. Financial support is gratefully acknowledged from Ministero dell'Università e della Ricerca Scientifica (MIUR) through the PRIN 2015 Grant 2015WTW7J3, from Compagnia di San Paolo in the framework of Project ID ROL 9361 and from MAECI in the framework of the Italy-Egypt bilateral protocol. M.T. is financially supported by the German Science Foundation (DFG) via the collaborative research center SFB 1249 and acknowledges support from Heidelberg Graduate School for Physics (HGSFP).

# Supporting Information

# Self-Organized Nanorod Arrays for Large-Area Surface-Enhanced Infrared Absorption

M. C. Giordano[a,], M. Tzschoppe[b], M. Barelli[a], J. Vogt[b], C. Huck[b], F. Canepa[a], A. Pucci[b,*],

and F. Buatier de Mongeot[a,*]

High sensitivity in Surface Enhanced Infrared Absorption (SEIRA) spectroscopy is demonstrated in the main manuscript by exploiting self-organized templates based on arrays of Au nanorod antennas.

The optical response and the morphological characterization of these metallic arrays showing dichroic plasmonic properties over a broadband spectrum are here described in detail in order to complete the information given in the main text.

The optical response of the Au nanorod arrays presented in the SEM image of Figure 2d of the manuscript is shown in Figure SI1 for a broadband spectral region ranging from the Near-Ultra Violet (NUV) to the Mid-Infrared (IR). A strong optical dichroism is detected at normal incidence transmittance when comparing the spectra acquired for longitudinal ($T_{\parallel}$ - continuous line) and transversal ($T_{\perp}$ - dashed line) polarization of the incident light with respect to the nanorods long axis. The $T_{\perp}$ spectrum shows a minimum tuned in the Visible (VIS) at about 20000 cm$^{-1}$ due to localized surface plasmon resonance excitation along the transversal axis of the rods while a broadband quasi-transparent optical behavior is detected over a broadband spectral region extending from the VIS to the Mid-IR. For longitudinal polarization (continuous red line in Fig.

SI1) we observe the characteristic maximum at 19000 cm$^{-1}$ due to inter-band transitions in Au at higher frequency while the transmittance gradually decreases in the VIS and Near-IR portion of the spectrum. Notably a broadband minimum is observed across the Near- and the Mid-IR spectrum due to plasmonic excitation parallel to the nanorod long axis.

Remarkably, a strong signal from the molecules is selectively detected (Fig. SI1 b) for longitudinal polarization (red continuous line), once a self-assembled monolayer of Octadecanthiol (ODT) is deposited on the surface. In parallel a very weak ODT signal is observed for transversal polarization (dashed line) which can be attributed to detached Au nanorods during dip-coating in solution (see SEM image acquired after ODT deposition in Fig. SI2). In order to extract the vibrational signal detected for transversal polarization the $T_\perp$ spectrum has been normalized as $T_\parallel / T_\perp$ (blue curve on Fig. SI1 b). The background optical contribution given by the plasmonic template has been fitted and subtracted from the curve, thus obtaining the vibrational signal shown in Fig. 3b. The vibrational signal shown in Fig. 3d, which is referred to a different sample characterized by higher density of antennas, has been quantified following the same procedure.

In Fig. SI 3 a, b we show representative SEM images of sample B and sample C, respectively. By means of the open source ImageJ software [1] we calculated the average antennas density per unit area at the optical excitation coordinates, which respectively read 37 μm$^{-2}$ for sample B and 44 μm$^{-2}$ for sample C. To have a more realistic estimate of the active sites effectively contributing to the SEIRA signal, as already described in the main manuscript, we also calculated the same quantity by taking into account only the antennas with length $L \geq 200$ nm. In this case the antennas density reads 12 μm$^{-2}$ for sample B and 18 μm$^{-2}$ for sample C.

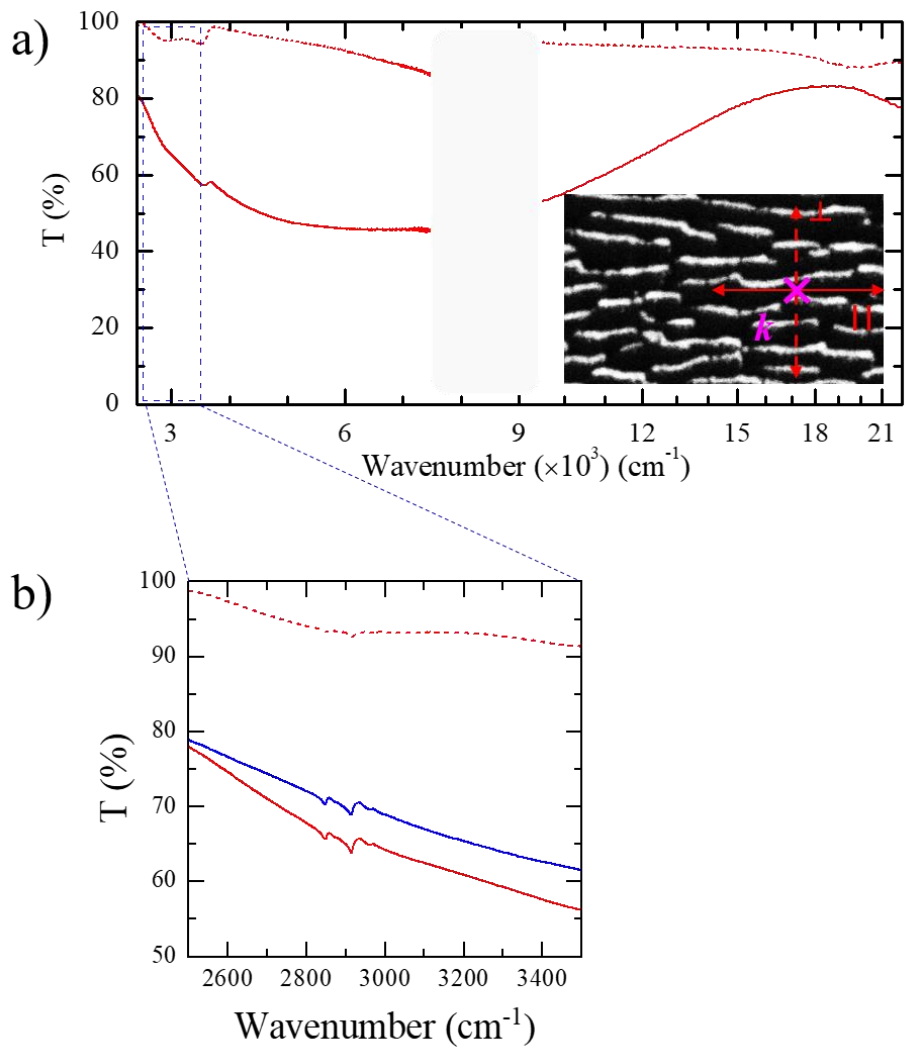

**Figure SI1 a)** Near-Ultra Violet, Visible and Near-/Mid-Infra-Red (IR) optical transmission spectra of Au nanorod arrays supported on glass templates. The spectra refer to the bare Au nanorods sample shown in Fig. 2 d for longitudinal ($T_{\parallel}$ - continuous line) and transversal ($T_{\perp}$ - dashed line) polarization of the excitation with respect to the nanorod long axis. **b)** Mid-IR transmittance spectra of the Au nanorod arrays coated with a monolayer of ODT molecules. The spectra $T_{\parallel}$ (continuous red line), $T_{\perp}$ (dashed red line) and the normalized signal $T_{\parallel} / T_{\perp}$ (blue line) are shown.

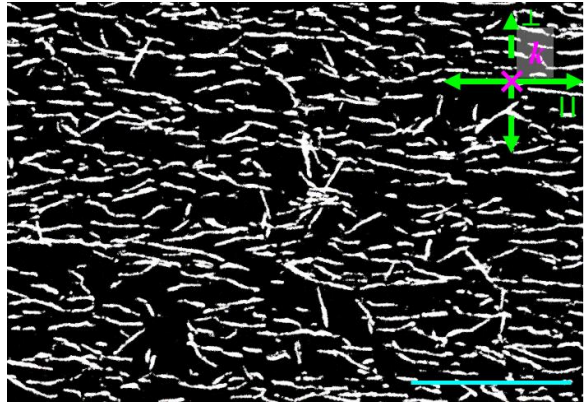

**Figure SI2** SEM image of Au nanorod arrays acquired after the dip-coating with the ODT solution. The scale bar corresponds to 1 µm. Due to the absence of a Ti adhesion layer between the Au nanorods and the glass template, some nanorods detached and redeposited with random orientation.

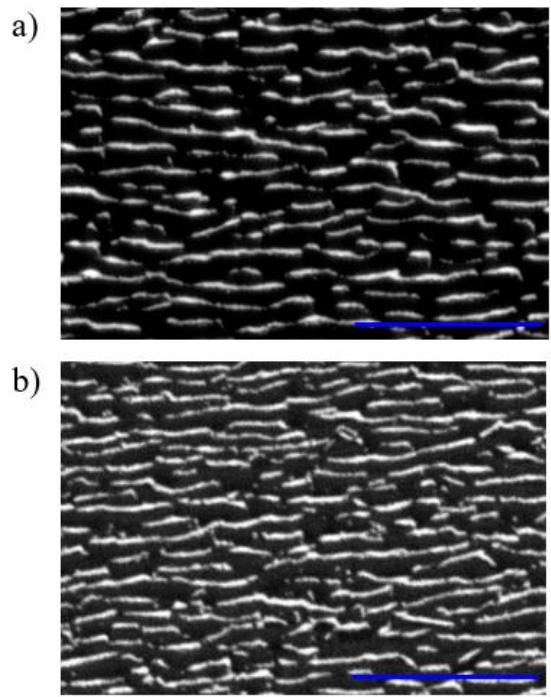

**Figure SI3.** SEM images acquired in the backscattered channel with a primary beam energy of 15 kV for sample B (a) and sample C (b). The blue scale bars correspond to 1 µm.

**Fit of the vibrational modes by Fano theory**

The asymmetric line shape of each vibrational band can be described by the Fano theory [2] which allows to perform the fit using the extinction line shape function $f(\varepsilon) = (q + \varepsilon)^2/(1 + \varepsilon^2)$, with $\varepsilon = 2(\omega - \omega_{\text{vib}})/\Gamma$, the resonance frequency of the vibrational excitation $\omega_{\text{vib}}$, the asymmetry parameter $q$, and the vibrational linewidth $\Gamma$. To apply the formula to the measured vibrational signal, we further normalized the function introducing the intensity $I$ and shifted its baseline. The formula used to describe our experimental data finally reads:

$$f(\omega) = 1 - \sum_{i=1}^{4} \frac{I_i}{q_i^2+1} \left( \frac{\left(2(\omega-\omega_{\text{vib},i})+q_i\Gamma_i\right)^2}{4(\omega-\omega_{\text{vib},i})^2+\Gamma_i^2} - 1 \right). \quad \text{(Eq. SI1)}$$

The sum accounts for the four different vibrational excitations observed in the experiment. In total, the formula contains 16 parameters (four parameters for each vibrational excitation). Eight of these parameters, namely the four vibrational frequencies and the four linewidths are well known. To further reduce the amount of parameters, we assume the same value of $q$ for all four excitations. This assumption is reasonable since all excitations are within a very small spectral range, much smaller than the linewidth of the plasmonic excitation. Finally, the experimental data has been fitted with five fit parameters (the asymmetry parameter $q$ and four intensities). The resultant fit, which nicely reproduces the measured spectrum, is shown in Figure SI4; the fit parameters are summarized in Table SI1. Asymmetry parameters around $q = \pm 1$ indicate a phase difference of about 90° between the narrow vibrational oscillator and the very broad plasmonic one. Details on how this can be calculated are exemplarily explained in Reference [3].

| Excitation | $\omega_{\text{vib}}$ | $\Gamma$ | $I$ | $q$ |
|---|---|---|---|---|
| Symmetric stretching vibration $CH_2$ | 2850* | 10.6* | 0.0318 | -1.32 |
| Symmetric stretching vibration $CH_3$ | 2877* | 9.9* | 0.00678 | |
| Asymmetric stretching vibration $CH_2$ | 2918* | 17.7* | 0.05726 | |
| Aymmetric stretching vibration $CH_3$ | 2960* | 15.2* | 0.00621 | |

**Table SI1.** Summary of the fit parameters used to describe the experimental data with four Fano-type excitations (Eq. 1) * Parameters for an ODT monolayer on gold, values taken from Ref. 4.

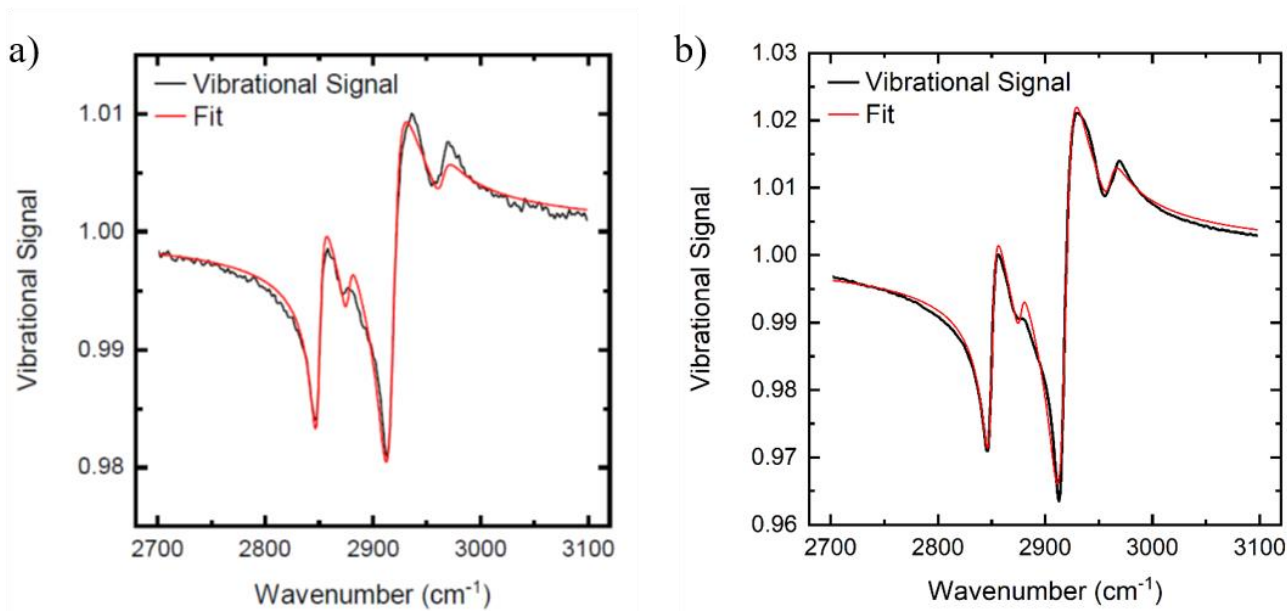

**Figure SI4.** Vibrational signal of an ODT monolayer adsorbed on the SO array of sample B (a) and sample C (b). The vibrational signal consists of four different excitations, which can be attributed to the symmetric and asymmetric stretching vibration of the $CH_2$ and $CH_3$ groups, respectively. The coupling with the plasmonic excitation results in asymmetric line shapes, which can be modeled by the Fano theory.

**Numerical Simulations**

Finite-difference time-domain (FDTD) simulations has been carried out by using the commercial software Lumerical FDTD solutions on the high performance cluster bwForCluster MLS&WISO (Production) as we already reported in several other contributions (see References [5-7] for more details), using the total-field scattered-field approach and perfectly matched layer boundary conditions in a distance of at least one wavelength away for all directions in space (for the simulation of a single antenna). For simulating arrays, we used periodic boundary conditions along the substrate's surface plane. To reduce computational cost, we took the advantage of the system's symmetry in the surface plane directions. From AFM data (Fig. 2 a-c), we roughly estimated the antenna geometry as half cylinder with a width of about 60 nm and an array periodicity of 200 nm in x- and y-direction [5]. The glass substrate was modelled as a dispersionless material with constant refractive index of $n = 1.50$ (within the modelled spectral region of (2000 - 6000) cm$^{-1}$). For gold, the dielectric function was described by a Drude model with a dielectric background of $\varepsilon_\infty = 7.0$, a plasma frequency of $\omega_p = 67900$ cm$^{-1}$, and an electronic scattering rate of $\omega_\tau = 384$ cm$^{-1}$ [8].

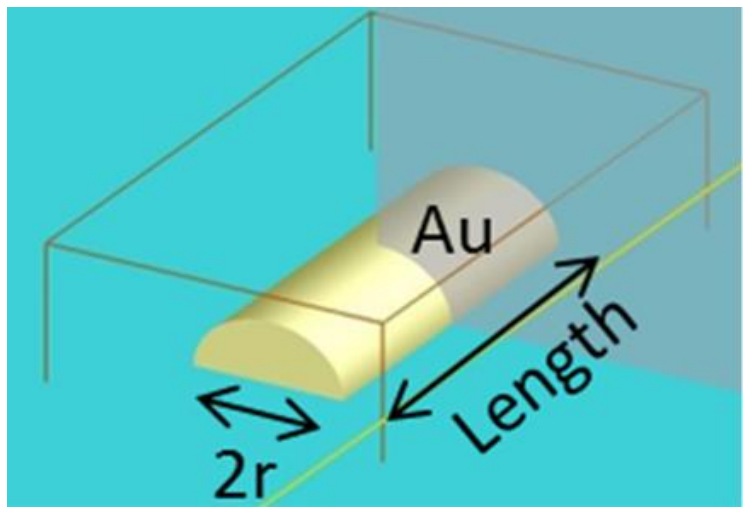

**Figure SI5.** Sketch of the antenna geometry which was used for the numerical simulations. The width was fixed to $w_{\text{antenna}} = 2r = 60$ nm, whereas the length was varied in order to determine the antenna length which corresponds to a plasmonic resonance in the desired spectral range.